\documentclass[dvips]{arxstspdf}
\usepackage{flushend}

\volume{23}
\issue{3}
\pubyear{2008}
\firstpage{318}
\lastpage{320}
\doi{10.1214/08-STS244C}
\referstodoi{10.1214/07-STS244}

\begin{document}
\begin{frontmatter}

\title{Comment: Quantifying the Fraction of Missing Information for Hypothesis
Testing in Statistical and Genetic Studies}
\runtitle{Comment}

\begin{aug}
\author[a]{\fnms{I-Shou} \snm{Chang}\ead[label=e1]{ischang@nhri.org.tw}},
\author[b]{\fnms{Chung-Hsing} \snm{Chen}\ead[label=e2]{chchen@nhri.org.tw}},
\author[c]{\fnms{Li-Chu} \snm{Chien}\ead[label=e3]{lcchien@nhri.org.tw}} \and
\author[d]{\fnms{Chao A.} \snm{Hsiung}\corref{}\ead[label=e4]{hsiung@nhri.org.tw}}
\runauthor{I.-S. Chang, C.-H. Chen, L.-C. Chien and C. A. Hsiung}

\affiliation{National Health Research Institutes,
National Central University,
National Health Research Institutes and
National Health Research Institutes}

\address[a]{I-Shou Chang is Investigator, Institute of Cancer Research and
Division of Biostatistics and Bioinformatics, National Health
Research Institutes, 35 Keyan Road, Zhunan Town, Miaoli County
350, Taiwan \printead{e1}.}
\address[b]{Chung-Hsing Chen is Postdoctor, Division of
Biostatistics and Bioinformatics, National Health Research Institutes,
35 Keyan Road, Zhunan Town, Miaoli Country 350, Taiwan \printead{e2}.}
\address[c]{Li-Chu Chien is Postdoctor, Division of Biostatistics and
Bioinformatics, National Health Research Institutes, 35 Keyan
Road, Zhunan Town, Miaoli County 350, Taiwan \printead{e3}.}
\address[d]{Chao A. Hsiung is Director, Division of Biostatistics and
Bioinformatics, National Health Research Institutes, 35 Keyan
Road, Zhunan Town, Miaoli County 350, Taiwan \printead{e4}.}

\end{aug}



\end{frontmatter}

Nicolae, Meng and Kong are to be congratulated on having treated an
important practical problem in many scientific inquiries in which
the investigator has chosen the testing procedure, but needs to
know the impact of the missing data on the test in terms of the
relative loss of information. To measure the relative information,
they propose to compare how the observed-data likelihood deviates
from flatness relative to the same deviation in the complete-data
likelihood. Several measures of this deviation expressed by
Bayesian method are explored and applied to the study of genetics
and genomics. As noted in their paper, these measures are
especially needed in small-sample problems with incomplete data.

We would like to explore the use of this type of measure in two
examples to indicate its wide applicability and some computational
issues. One concerns infectious disease data, which are usually
highly dependent and incomplete; the investigators often need to
decide if more data are needed, and in case they are, to know the type
of data that is most desirable. The other concerns a test on the
shape of a regression function; we will apply the Bayesian measure
of relative information to select design points for collecting
more data.

Because Bayesian tests are more tractable and natural than a
frequentist approach in these two examples, we consider the
following extensions of their (25) for the measure of relative
information:
\def\theequation{BI\arabic{equation}}
\setcounter{equation}{2}
\begin{eqnarray}
\qquad \hspace*{5pt}
&& \mathrm{E}_0\{ \operatorname{Var}[ \operatorname{lod}
( \theta_0,\theta | Y_{\mathrm{ob}} ) | Y_{\mathrm{ob}} ] \}
\nonumber \\
&&\quad{} \cdot
\biggl(\mathrm{E}_0\biggl\{ \operatorname{Var}
[ \operatorname{lod}( \theta_0,\theta | Y_{\mathrm{ob}} ) | Y_{\mathrm{ob}} ]
\label{ebi3} \\
&&\hspace*{41pt}{}
+ \operatorname{Var}\biggl[\log {P(Y_{\mathrm{co}} | Y_{\mathrm{ob}},\theta)
\over{P(Y_{\mathrm{co}} | Y_{\mathrm{ob}},\theta_0)}}\bigg| Y_{\mathrm{ob}} \biggr]
\biggr\}\biggr)^{-1}\hspace*{-5pt}
\nonumber \\
&& \mathrm{E}_0 \biggl\{\operatorname{Var}[ \operatorname{lod}( \theta_0,\theta |
Y_{\mathrm{ob}} ) | Y_{\mathrm{ob}} ]
\nonumber\\
&&\hspace*{16pt}{} \cdot \biggl( \operatorname{Var}
[ \operatorname{lod}( \theta_0,\theta | Y_{\mathrm{ob}} ) | Y_{\mathrm{ob}} ]
\label{ebi4} \\
&&\hspace*{29pt}{}
+ \operatorname{Var}\biggl[\log {P(Y_{\mathrm{co}} | Y_{\mathrm{ob}},\theta)
\over{P(Y_{\mathrm{co}} | Y_{\mathrm{ob}},\theta_0)}}\bigg| Y_{\mathrm{ob}} \biggr]  \biggr)^{-1} \biggr\} .
\nonumber
\end{eqnarray}
Here $\mathrm{E}_0$ means average over $\theta_0$ from the
conditional posterior distribution on the null hypothesis. To
shorten the presentation, we use only (\ref{ebi3}) in the following
discussion.

\section{Infectious Disease Data}

As discussed in
Rhodes, Halloran and Longini
(\citeyear{RhodesHalloranLongini96}), there are several levels of
information in the study of infectious disease data and it is of
interest to decide the level of information in the study. We
consider two levels of information in a simple model to illustrate
the way that (\ref{ebi3}) may be used in this situation. Suppose there is
a collection of disjoint households that suffer a transmissible
disease and an individual can only be infected by members in the
same household. We assume an S--I--R model; at any time point, each
individual is in one of the three states: susceptible (S),
infectious (I) or removed (R); a susceptible individual may become
infectious and an infectious individual may become removed. Assume
there are $m$ people in one household. The transition of the
health status of people in one household is described by the
following counting process. We note that counting process
modeling of infectious disease data is discussed in
Becker (\citeyear{Becker89}) and
Andersson and Britton (\citeyear{ABritton2000}), among others.

For $i=1,\ldots,m$, let $N_i(t)$ be 1 if the $i$th individual has
been infected at time $t$ and be 0 if not; for $i=m+1,\ldots,2m$,
let $N_i(t)$ be 1 if the $(i-m)$th individual has been removed at time
$t$ and be 0 if not. Let $I(t)$ denote the number of infectious
people at time $t$. Here $t\geq0$. Assume $N_1(0)=1$, which means
this individual is the first infected person. Assume that
$P(N_i(t+h)-N_i(t)|{\mathcal F}_t)=h\lambda_i(t)+o(h)$. Here
$\lambda_i(t)=\beta_0 \exp (\beta_1Z_i)I(t-)(1-N_i(t-))$
for $i=1,\ldots,m$, and
$\lambda_i(t)=\gamma_0(N_{i-m}(t-)-N_i(t-))$ for
$i=m+1,\ldots,2m$; ${\mathcal F}_t$ is the history up to time $t$.
The parameters $\beta_0$ and $\gamma_0$ are respectively called
the infection rate and the removal rate.

Assuming the covariate $Z_i$ has value 0 or 1, we are interested
in testing the hypothesis $H_0$ that $\beta_1$ is less than 0.
When $Z_i=1$ means that the $i$th individual has been vaccinated,
$\beta_1$ may represent the efficacy of the vaccine.

We assume the removal times of all the removed individuals are
observable and their infection times are not observable except the
first one in the household, which is assumed to be zero. We note
that it is often easier to obtain removal times than infection
times; the latter are often hard, if not impossible, to get; the
sole purpose of assuming that the first infection time is
observable is to simplify the presentation.

Suppose we have collected the observed data and decided to test
the hypothesis $H_0$ by considering the ratio of the posterior
probability to the prior probability of the event [$\beta_1<0$].

Viewing all the infection times except the first one in each
household as missing data, we can use (\ref{ebi3}) to measure the
fraction of missing information. Alternatively, we may consider
the removal times of additional four, say, households as missing
data and calculate its (\ref{ebi3}). These two (\ref{ebi3})s might be useful in
deciding, when additional data are needed, which type of additional
data is more desirable. We illustrate this method in the following
simulation studies.

Assuming $\beta_0=1$, $\beta_1=-0.5$, $\gamma_0=1$, there are 6
members in each household and there are 20~households, we generate
a set of observed data; assuming the priors for $\beta_0$ and
$\gamma_0$ are exponentially distributed as $\operatorname{Exp}(1)$ and that
for $\beta_1$ is standard normal, we use MCMC to generate the
posterior distributions of the parameters.

The relative information (\ref{ebi3}) has values 0.795 and 0.288,
respectively, for the missing data being infection times and for
that being additional four household removal times. This seems to
suggest that obtaining additional four household removal times is
more desirable for this set of observed data. By the way, the
prior probability of [$\beta_1 <0$] is 0.5 and the posterior
probability of [$\beta_1 <0$], given the removal times of the
20~households, is 0.739. Although we have treated only an
oversimplified example, this simulation study seems to suggest
that the relative information measure proposed by
Nicolae, Meng and Kong (\citeyear{nicolea08})
is useful in infectious disease data analysis.

\section{A Test for Monotonicity of a Regression Function}

Let $\mathcal{S}$ denote the set of all continuous functions on
$[0,1]$ and $\mathcal{I}$ denote the set of all nondecreasing
continuous functions on $[0,1]$. Consider the regression model
\[
Y_{k}=F(X_k)+\sigma\varepsilon_k,
\]
for some $F$ in $\mathcal S$. Here for $k=0,\ldots,K$, $Y_{k}$ is
a response variable, $X_k$ is a constant design point in $[0,1]$,
and the errors $\{\varepsilon_k \}$ are assumed to be independent
and standard normal; $\sigma$ is a positive constant.

We are interested in testing the hypothesis $H_0$ that the
regression function $F$ is nondecreasing and wish to know the way
to collect more data properly. We will introduce a probability
measure on $\mathcal S$, and consider a Bayesian approach.

Let $\mathcal{B}=\bigcup_{n=1}^{\infty} ( \{n\} \times \mathbb{R}^{n+1})$ and
$\varphi_{i,n}(t)=C_{i}^{n}t^{i}(1-t)^{n-i}$ for $t\in[0,1]$. For
$b_{n}=(b_{0,n},\ldots,b_{n,n})$, we define $F_{b_{n}}(t)=
F_{b_{n}}(n,b_{0,n},\ldots,b_{n,n},t)=\sum_{i=0}^{n}b_{i,n}\varphi_{i,n}(t)$.
We note $F_{b_{n}}$ is called a Bernstein polynomial with
coefficients $b_{0,n},\ldots,b_{n,n}$. It is readily seen that
$F_{b_{n}}(\cdot)$ is a member of $\mathcal{S}$ and it is a member
of $\mathcal{I}$, if $b_n\in\{b_n | b_{0,n}\leq\cdots\leq b_{n,n}\}$.
Let $S_{n}=\{ F_{b_{n}} |  b_{n}\in\mathbb{R}^{n+1}\}$.
It is clear that $\mathcal S\supset\bigcup_{n=1}^{\infty}S_{n}$.
A probability measure $\pi$ can be introduced on $\mathcal{S}$
as follows. Let $\pi_{n}$ be a conditional density on
$\mathbb{R}^{n+1}$ and $p$ a probability mass function on
$\{1,2,\ldots\}$; define $\pi(n,b_{n})=p(n)\pi_{n}(b_{n})$, which
introduces a probability measure on
$\bigcup_{n=1}^{\infty} (\{n\} \times\mathbb{R}^{n+1} )$. Identifying a Bernstein
polynomial with its order and coefficients, we can regard $\pi$ as
a probability on $\bigcup_{n=1}^{\infty} S_{n}$, hence on
$\mathcal S$. Priors of this form are referred to as Bernstein
priors.

Chang et al. (\citeyear{Changetal2007}) showed that suitably introduced Bernstein
priors facilitate the estimation of $F$ under various shape
restrictions. In fact, this approach also provides a direct
assessment of the hypothesis $H_0$ that $F$ is in $\mathcal I$ by
considering the ratio of the posterior probability to the prior
probability of the set $\mathcal I$. We note that the Bernstein
priors used in
Chang et al. (\citeyear{Changetal2005}) and
Chang et al. (\citeyear{Changetal2007}) have
large supports and, yet, take into consideration the shape
restrictions, and the prior on $\mathcal S$ that we use in the
following simulation is constructed in the spirit of these
references and motivated by the simple observation that if
$b_{i,n}$ is in [$\tau_1,\tau_2$] for every $i$, then $F_{b_n}$ is
in [$\tau_1,\tau_2$], and a continuous function with values in
[$\tau_1,\tau_2$] can be approximated by Bernstein polynomials
with coefficients contained in [$\tau_1,\tau_2$].

Suppose we have collected response variables at $X_0,\ldots,X_K$
and would like to know the relative\break information of the observed
data when more\break response variables are taken at additional design
points $x_0,\ldots,x_L$. The following simulation studies are
meant to illustrate the use of (\ref{ebi3}) in this problem. Assume
$F(t)=0.6t$ for $t$ in $[ 0, 1 ]$ and $\sigma=0.4$. Let $K=9$ and
$X_k=k/9$ for $k=0,\ldots,9$. We generate one set of data
according to this specification, and then calculate (\ref{ebi3}) under
several missing data scenarios. When $L=K$ and
$x_0=X_0,\ldots,x_L=X_L$, we find (\ref{ebi3}) is equal to 0.139. When
($0,x_0,\ldots,x_4,\break 0.5$) form an equal length partition of the
interval $[ 0, 0.5 ]$ and ($0.5,x_5,\ldots,x_9,1$) form an equal
length partition of the interval $[ 0.5, 1 ]$, we find (\ref{ebi3}) is
equal 0.346. This shows that the former design points would be
preferable to the latter when additional data are needed.

To have some idea for the case $L=2K$, we find (\ref{ebi3}) is 0.052 if
$x_{2k}=x_{2k+1}=X_k$ for $k=0,\ldots,K$, and is 0.217 if
($0,x_0,\ldots,x_9,0.5$) form an equal length partition of the
interval $[ 0, 0.5 ]$ and ($0.5,x_{10},\ldots,\break  x_{19},1$) form an
equal length partition of interval $[ 0.5, 1 ]$. We note that the
prior probability of $\mathcal I$ is 0.0006 and the posterior
probability of $\mathcal I$ is 0.0015. In summary, we find the
measure of relative information (\ref{ebi3}) useful in selecting extra
design points for data collection in this regression example.

\section{Some Computational Remarks}

Nicolae, Meng and Kong (\citeyear{nicolea08}) pointed out that~(24) may be problematic
because of the large variability in the likelihood ratios. That
this problem does appear in the above two examples is the sole
reason that only extensions of (25) are used here.

Because we work with Bayesian tests, in which there are already
specified priors, it seems natural to use the corresponding
posteriors in the calculation of (24) and~(25) and their
extensions like (\ref{ebi3}) and (\ref{ebi4}). In particular, the $\mathrm{E}_0$ in
(\ref{ebi3}) and (\ref{ebi4}) is the conditional posterior probability on the
null hypothesis. It may happen that the (unconditional) posterior
probability of the null hypothesis is so small that sampling from
the conditional posterior probability needs large computation
time, which may make the calculation of (\ref{ebi3}) hard. In this
connection, we would like to note that although the posterior
probability for the above regression problem is somewhat small, it
is still manageable.

\section*{Acknowledgment}

This work was supported in part by NSC Grant NSC
96-2118-M-400-002-MY2.

\end{document}